# Anisotropic Superconducting Properties of $MgB_2$ Single Crystals.

Yu. Eltsev, S. Lee, K. Nakao, N. Chikumoto, S. Tajima, N. Koshizuka, and M. Murakami
Superconductivity Research Laboratory, ISTEC, 10-13 Shinonome, 1-chome, Koto-ku, Tokyo, 135-0062, Japan

### *Abstract*

*In-plane electrical transport properties of $MgB_2$ single crystals grown under high pressure of 4-6 GPa and temperature of 1400-1700°C in Mg-B-N system have been measured. For all specimens we found sharp superconducting transition around 38.1-38.3K with $\Delta T_c$(10-90%) within 0.2-0.3K. Estimated resistivity value at 40K is about $1\mu\Omega cm$ and resistivity ratio $\rho(273K)/\rho(40K)=4.9\pm0.3$. Results of measurements in magnetic field up to 5.5T perpendicular to Mg and B planes ($H_\perp$) and up to 9T in parallel orientation ($H_{//}$) show temperature dependent anisotropy of the upper critical field with anisotropy ratio $\gamma=H_{c2//}/H_{c2\perp}$ increasing from 2.2 close to $T_c$ up to about 3 below 30K. Strong deviation of the angular dependence of $H_{c2}$ from anisotropic mass model has been also found.*

## Introduction

Recent discovery of superconductivity at T=39K in magnesium diboride [1] attracted great attention of numerous research groups. Transport and magnetization measurements performed on sintered polycrystalline samples of $MgB_2$ revealed properties typical for a type-II superconductor [2-8]. Important role of phonons in superconducting interaction in $MgB_2$ was suggested from observation of the boron isotope effect [9] as well as from theoretical calculations of $MgB_2$ band structure [10]. Clear signs of anisotropic behavior of $MgB_2$ have been observed in measurements performed on thin films [11,12], aligned crystallites [13] and fine powder [14] samples. Recently Lee *et al.* reported on growth of $MgB_2$ single crystals [15]. Availability of single crystals gives a nice opportunity for direct probe of the electronic anisotropy of this compound. Here we present the results of the detailed study of the anisotropic properties of $MgB_2$ single crystals probed by the in-plane electrical transport measurements.

## Experimental

Magnesium diboride single crystals have been grown in quasi-ternary Mg-$MgB_2$-BN system at high pressure and temperature of 4-6 GPa and 1400-1700$^o$C respectively. The $MgB_2$ precursor was prepared from magnesium powder (99.9% Rare Metallic Co.) and amorphous boron (97% Hermann C.Starck). Single crystal growth was performed in BN crucibles in the presence of longitudinal temperature gradient of about 200$^o$C/cm in a cubic-anvil press (TRY Engineering). In optimal conditions shiny gold-coloured single crystals of size up to 0.7mm were grown. Several plate-like single crystals of size of about 0.5x0.1x0.05$mm^3$ have been chosen for our study. The in-plane transport measurements have been performed in a usual four probes linear geometry with transport current directed along Mg and B planes. Electrical contacts were made using gold or silver paste without subsequent heat treatment. Contact resistance was around 1 for current contacts and slightly higher (about 3-5 ) for potential ones. To measure current-voltage response, we used usual low frequency (17Hz) lock-in ac technique with excitation current in the range 0.2-2.0mA and voltage resolution of about 0.3nV. To allow for accurate single crystal alignment with respect to magnetic field the samples were mounted in a rotatable sample holder with an angular resolution of 0.05$^o$.

**Results and discussion**

All the results obtained on different $MgB_2$ single crystals used in this study demonstrate remarkable reproducibility, thus, suggesting high quality of our samples. In particular, for all crystals we found sharp superconducting transition with $T_c$, defined as the resistivity onset, about 38.5K and transition width $\Delta T_c(10\%-90\%)<0.3$K. Typical zero field superconducting transition is shown in Fig. 1. Just above $T_c$ we estimate $\rho=1\pm0.15\mu\Omega$ cm and residual resistivity ratio RRR= $\rho(273K)/\rho(40K)=4.9\pm0.3$. In Fig. 2 we present angular dependence of resistivity with magnetic field rotated around two different axes ($\theta=0°$, 180° and 360° corresponds to H//c orientation). For both R($\theta$) curves one can see two sharp minima at $\theta=90°$ and 270° when magnetic field direction is parallel to ab-planes. This result gives additional proof of high quality of our samples since any deviation from the ideal crystal structure (e.g. small misorientation of different crystallites) would result in more complicated R($\theta$) dependence.

Fig. 3 shows resistive superconducting transitions at various magnetic fields up to 5.5T for magnetic field perpendicular to Mg and B sheets (upper panel) and up to H=9T for H//ab planes (lower panel). Much more rapid $T_c$ suppression by perpendicular field compared to parallel one clearly demonstrates the upper critical field anisotropy of $MgB_2$ compound. Using the data shown in Fig. 3 we constructed magnetic phase diagram (see Fig. 4). As mentioned above, we associate resistivity onset with the upper critical field, $H_{c2}(T)$, while resistive transition end-point may be roughly attributed to the irreversibility field, $H^*(T)$. At T>30K for both field orientations $H_{c2}(T)$ dependence shows distinct positive curvature. In the same temperature range $30K<T<T_c$ the upper critical field anisotropy ratio $\gamma=H_{c2//}/H_{c2\perp}$ displays remarkable temperature dependence increasing from $\gamma=2.3$ near $T_c$ up to $\gamma=3$ at T=30K. Below T=30K $H_{c2//}(T)$ as well as $H_{c2\perp}(T)$ exhibit linear increase with decreasing temperature and $\gamma$ remains nearly unchanged. From the available data we estimate $H_{c2\perp}(0)=7.0-7.5$T and $H_{c2//}(0)=\gamma H_{c2\perp}(0)=21-22$T. According to the relations for anisotropic superconductors $H_{c2\perp}(T)=\Phi_0/(2\pi\xi_{ab}^2)$, where $\Phi_0$ is the flux quantum, and $\gamma=H_{c2//}(0)/H_{c2\perp}(0)=\xi_{ab}(0)/\xi_c(0)$ this corresponds to the in-plane and out-of-plane coherence length $\xi_{ab}(0)\approx 68$Å and $\xi_c(0)\approx 23$Å respectively.

We also note that anisotropic nature of $MgB_2$ material strongly affects its irreversible properties. From Fig. 3 and Fig. 4 one can see substantial broadening of superconducting transitions in perpendicular field orientation that reflects strong suppression of irreversibility field below $H_{c2}$: $H^*\approx 0.5H_{c2}$. In striking contrast, for parallel field orientation we find $H^*_{//} \approx 0.9 H_{c2//}$.

From Fig. 3 one can see unusually high noise level appearing at the lower part of superconducting transition at H>2T in perpendicular field orientation. Similar noise has been reproducibly observed in all $MgB_2$ crystals studied in this work suggesting its intrinsic origin. We postpone further discussion of this noisy behavior since it is outside the scope of present study.

Finally we present the results of measurements of the angular dependence of the upper critical field by sweeping magnetic field at constant temperature and at various sample orientations. In Fig. 5 we show the data for three different temperatures. Similar to results obtained from R(T) measurements, close to $T_c$ we get anisotropy ratio $\gamma = H_{c2//}/H_{c2\perp} = 2.3$. With the decrease of temperature $\gamma$ increases and at T=24K we obtain $\gamma = 3.15$. From Fig. 5 one can also see that $H_{c2}(\theta)$ dependence does not follow the theoretical expression $H_{c2}(\theta) = H_{c2//}/(\cos^2(\theta) + \gamma^2 \sin^2(\theta))^{-0.5}$ obtained within anisotropic mass model [16]. Furthermore, close to H//ab orientation experimental $H_{c2}(\theta)$ dependence clearly demonstrate cusp structure. It should be noted, that the shape of experimental $H_{c2}(\theta)$ dependence in the angle range $\theta = -45°+45°$ does not depend on the way how to determine $H_{c2}$ from R(T) curve. With any criterion ranging from $R(T_c)=0.5R_n$ up to $R(T_c)=0.98R_n$ deviation of experimental $H_{c2}(\theta)$ dependence from anisotropic mass model expression is evident. At present the origin of strong disagreement between theory and experiment is not clear since obtained value of the out-of-plane coherence length of about 23Å substantially exceeds interlayer spacing in $MgB_2$ and, thus, rules out a possibility of 2D behavior.

## Conclusions

We performed the in-plane transport measurements on $MgB_2$ single crystals to probe anisotropic behavior of this recently discovered superconductor. The results indicate moderate value of the upper critical field anisotropy ratio that is nearly temperature independent at about 3.0 below 30K and monotonously decreases to $\gamma = 2.3$ approaching $T_c$.

## Acknowledgements

This work is supported by the New Energy and Industrial Technology Development Organization (NEDO) as Collaborative Research and Development of Fundamental Technologies for Superconductivity Applications.

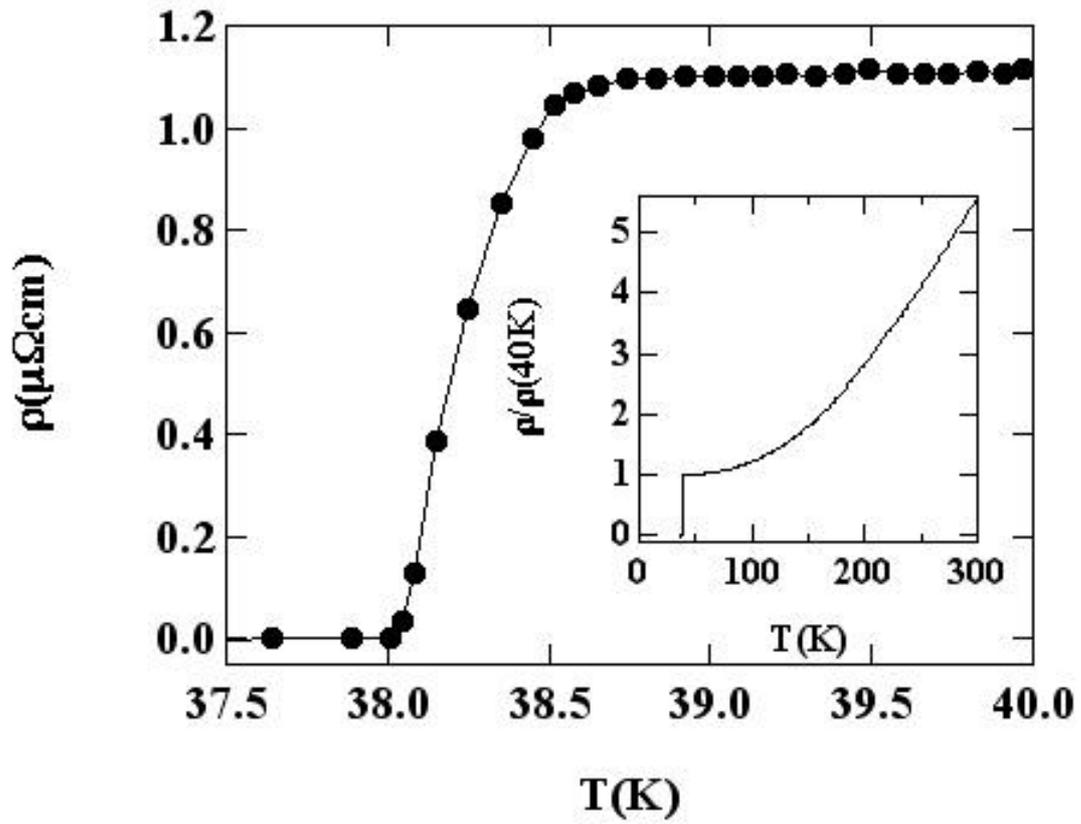

**Fig.1.** Typical zero-field superconducting transition for $MgB_2$ single crystal. Inset: Temperature dependence of the in-plane resistivity up to room temperature.

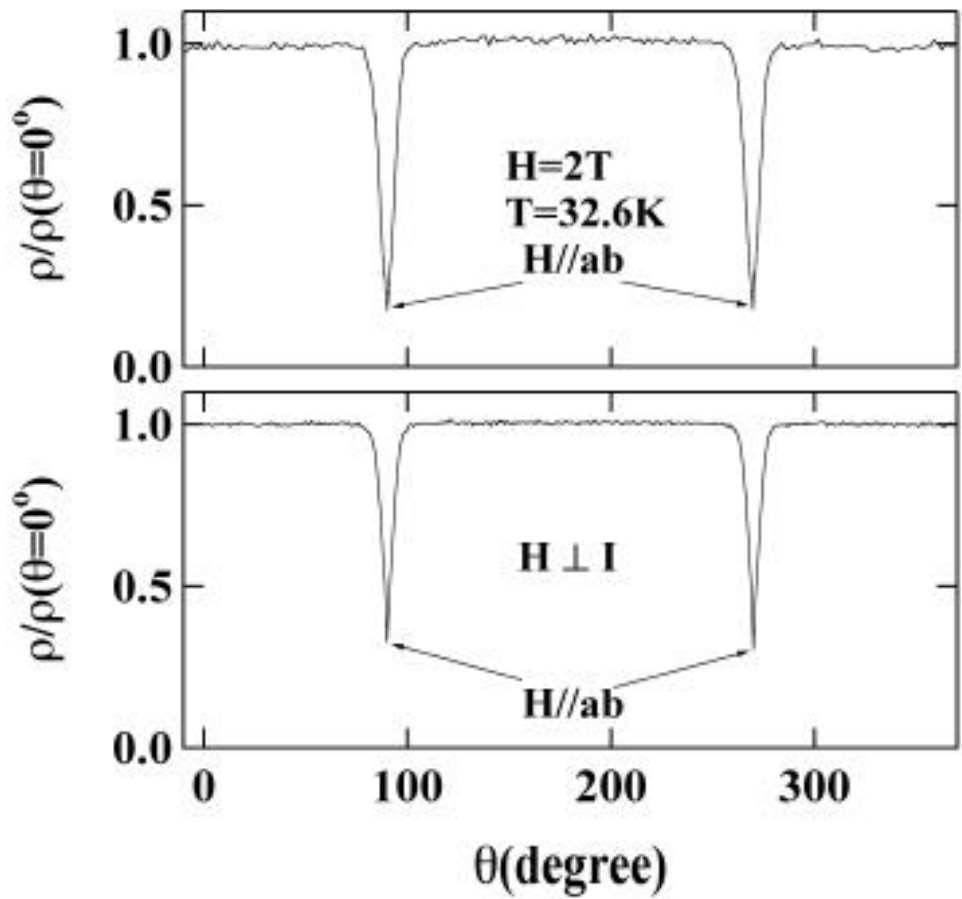

**Fig.2.** Angular dependence of the resistance as the field is rotated. =90º and =270º correspond to H//ab-planes. In the lower panel the current direction is perpendicular to magnetic field for all angles. In the upper panel the angle between current direction and magnetic field changes as angle between H and ab-planes.

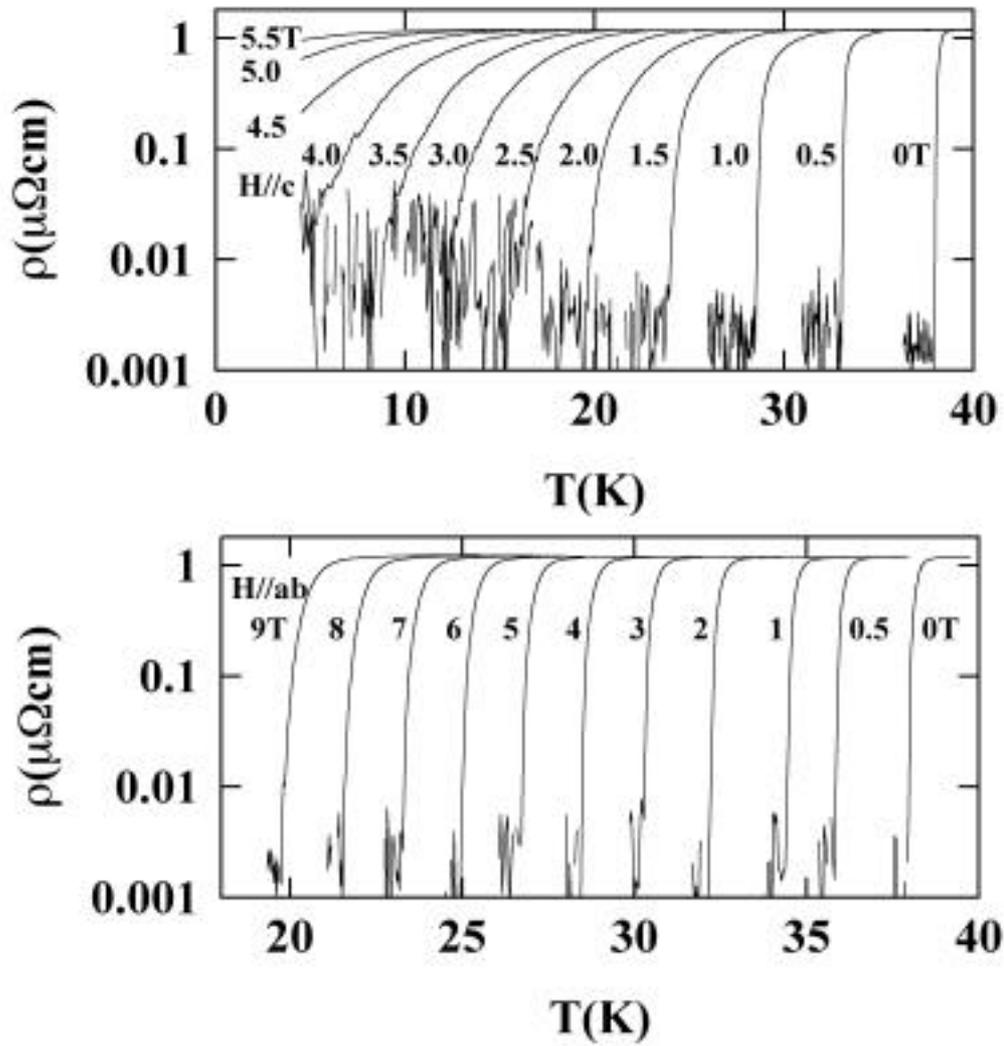

**Fig.3.** Upper panel: Superconducting transitions at various magnetic fields up 5.5T applied perpendicular to Mg and B planes. Lower panel: Superconducting transitions in parallel fields up to 9T. In both field orientations I=1mA and its direction is perpendicular to magnetic field. Different transition temperatures for a given applied field demonstrate the upper critical field anisotropy of $MgB_2$ single crystal.

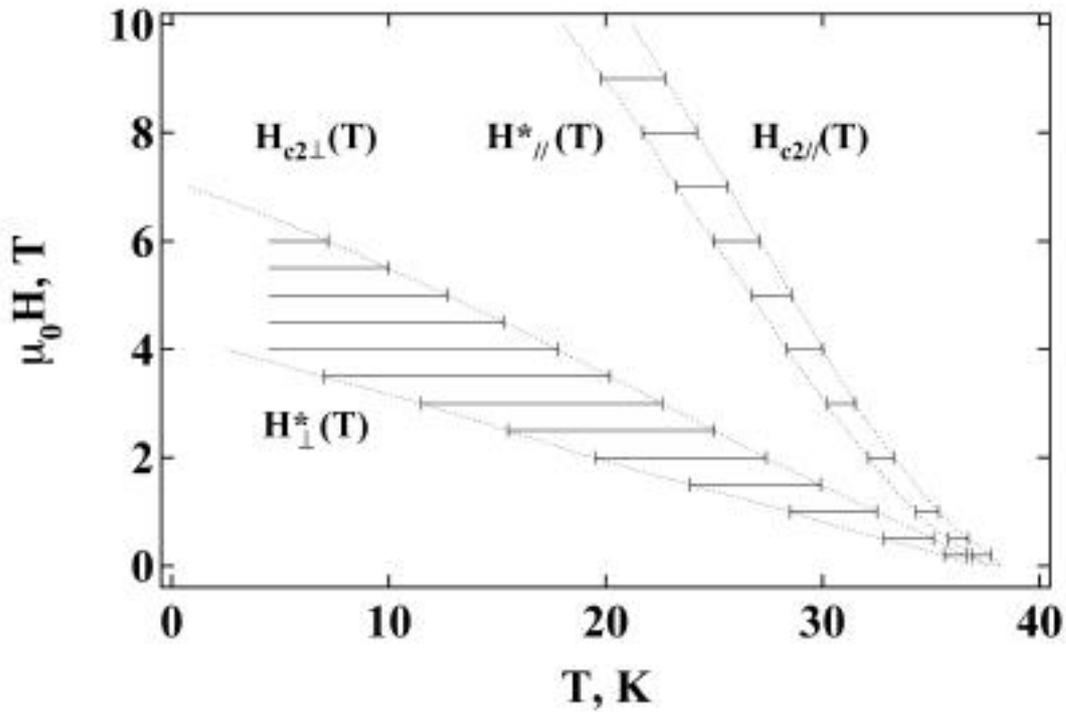

**Fig.4.** Magnetic phase diagram of $MgB_2$ single crystal deduced from the in-plane temperature-dependent resistivity data. Resistivity onset and transition end-point are vertical bars. For both samples the solid lines demonstrate transition width. The dashed lines are guides for eye and represent temperature dependence of the upper critical field ($H_{c2//}$ and $H_{c2\perp}$) and irreversibility field ($H^*_{//}$ and $H^*_\perp$) in parallel and perpendicular magnetic field orientation respectively.

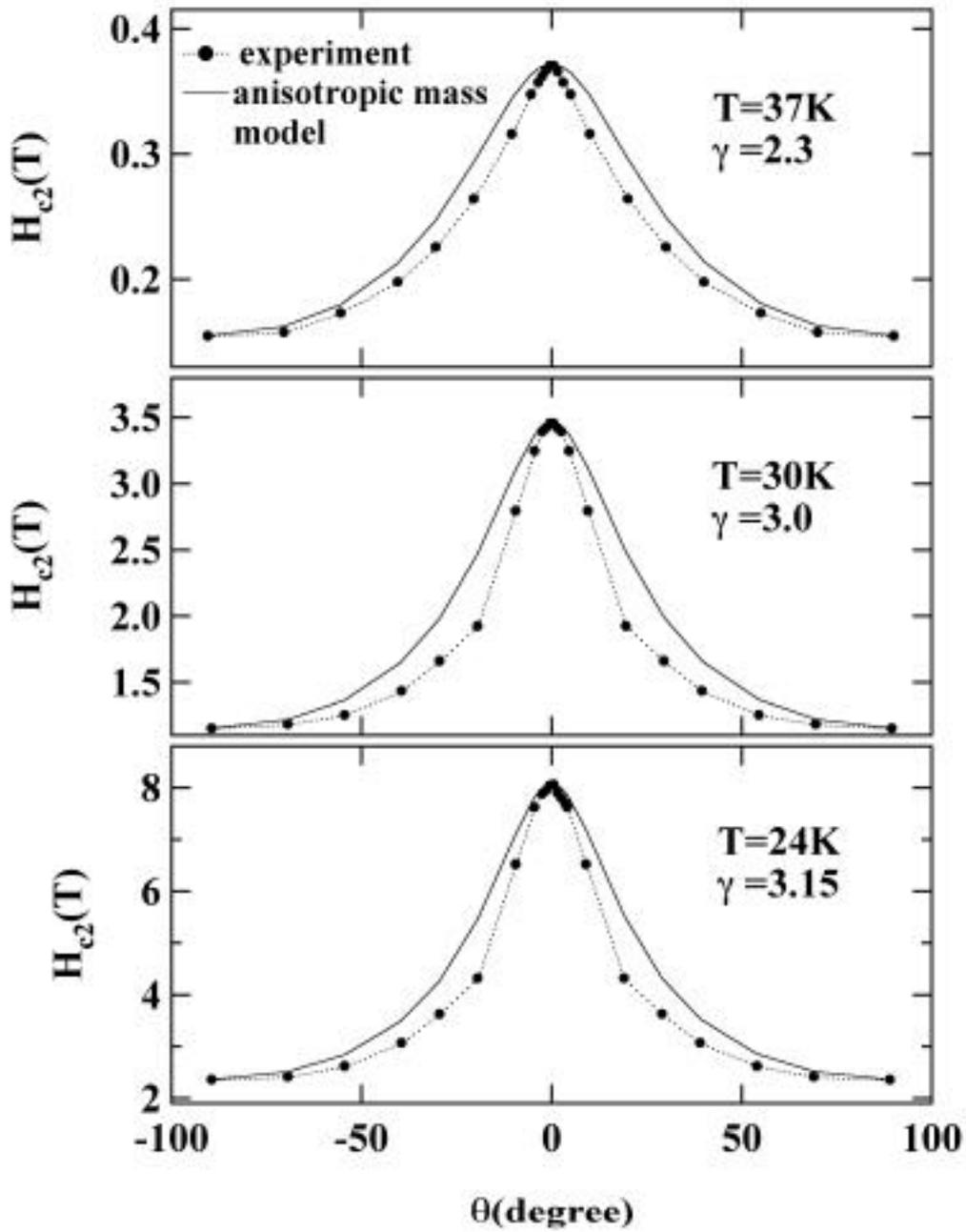

**Fig. 5** Angular dependence of the upper critical field at three different temperatures. The solid lines represent the angular dependence of the upper critical field according to anisotropic mass model. The dashed lines are guides for eye.